\documentclass[pra,twocolumn,showpacs]{revtex4}
\usepackage{amsmath}
\usepackage{amssymb}
\usepackage{epsfig}

\begin{document}

\title{ Stable and unstable vortex knots in a trapped Bose-Einstein condensate }
\author{V.~P. Ruban}
\email{ruban@itp.ac.ru}
\affiliation{L.D. Landau Institute for Theoretical Physics RAS, Moscow, Russia}
\date{\today}

\begin{abstract}
The dynamics of a quantum vortex torus knot ${\cal T}_{P,Q}$ and similar knots in an atomic
Bose-Einstein condensate at zero temperature in the Thomas-Fermi regime has been considered 
in the hydrodynamic approximation. The condensate has a spatially nonuniform equilibrium 
density profile $\rho(z,r)$ due to an external axisymmetric potential. It is assumed that 
$z_*=0$, $r_*=1$ is a maximum point for function $r\rho(z,r)$, with 
$\delta (r\rho)\approx-(\alpha-\epsilon) z^2/2 -(\alpha+\epsilon) (\delta r)^2/2$ at small 
$z$ and $\delta r$. Configuration of knot in the cylindrical coordinates is specified by 
a complex $2\pi P$-periodic function $A(\varphi,t)=Z(\varphi,t)+i [R(\varphi,t)-1]$. 
In the case  $|A|\ll 1$ the system is described by relatively simple approximate equations
for re-scaled functions $W_n(\varphi)\propto A(2\pi n+\varphi)$, where $n=0,\dots,P-1$, and
$iW_{n,t}=-(W_{n,\varphi\varphi}+\alpha W_n -\epsilon W_n^*)/2-\sum_{j\neq n}1/(W_n^*-W_j^*)$.
At $\epsilon=0$, numerical examples of stable solutions as 
$W_n=\theta_n(\varphi-\gamma t)\exp(-i\omega t)$ with non-trivial topology have been found for $P=3$.
Besides that, dynamics of various non-stationary knots with $P=3$ was simulated, and in some
cases a tendency towards a finite-time singularity has been detected. For $P=2$ at small 
$\epsilon\neq 0$, rotating around $z$ axis configurations of the form
$(W_0-W_1)\approx B_0\exp(i\zeta)+\epsilon C(B_0,\alpha)\exp(-i\zeta) + \epsilon D(B_0,\alpha)\exp(3i\zeta)$
have been investigated, where $B_0>0$ is an arbitrary constant, $\zeta=k_0\varphi -\Omega_0 t+\zeta_0$, 
$k_0=Q/2$, $\Omega_0=(k_0^2-\alpha)/2-2/B_0^2$. In the parameter space $(\alpha, B_0)$,
wide stability regions for such solutions have been found. In unstable bands,
a recurrence of the vortex knot to a weakly excited state has been noted to be possible.
\end{abstract}
\maketitle

\section{Introduction}

Dynamics and statics of quantum vortex filaments is among the most important and interesting problems 
in the theory of Bose-Einstein-condensed cold atomic gases  (see review \cite{F2009}, and references therein). 
The problem is greatly complicated (but simultaneously enriched) when the condensate is in an external 
trap potential $V({\bf r})$, since in that case the vortex motion occurs against a spatially nonuniform
density background. A large body of research was reported about this topic (see, e.g., 
\cite{SF2000,FS2001,R2001,AR2001,GP2001,A2002,AR2002,RBD2002,AD2003,AD2004,SR2004,D2005,
Kelvin_vaves,ring_istability,v-2015,reconn-2017,top-2017}), but the subject has not yet been exhausted.
In particular, dynamics of topologically nontrivial vortex configurations was considered so far
at a uniform density background only (see \cite{RSB999,MABR2010,POB2012,KI2013,POB2014,LMB016,KKI2016}, 
and references therein). To fill partly this gap in the theory, some simplest vortex knots in trapped 
axisymmetric condensates will be studied in the present work.

Let us first make necessary preliminary remarks that later will ensure quite simple derivation of 
approximate equations of motion for knotted vortex filaments, suitable for analysis.

In the general case, vortices essentially interact with potential excitations and with over-condensate
atoms. But if the condensate at zero temperature is in the so called Thomas-Fermi regime (a core width 
$\xi$ is much less than a typical vortex size $R_*$), then potential degrees of freedom can be neglected,
and the ``anelastic'' hydrodynamical approximation can be used (see appropriate examples in 
\cite{SF2000,FS2001,R2001,A2002,SR2004,BN2015,R2017-1,R2017-2}). At the formal level this means that
the condensate wave function $\Psi({\mathbf r},t)=|\Psi|\exp(i\Phi)$ is completely determined by
geometry of the vortex filament ${\mathbf R}(\beta,t)$ (where $\beta$ is an arbitrary longitudinal
parameter, $t$ is the time) and corresponds to a minimum of the Gross-Pitaevskii functional
(all the notations are standard)
\begin{equation}
{\cal H}=\int \Big[\frac{\hbar^2}{2m_{\rm at}}|\nabla\Psi|^2
+[V({\bf r})-\mu]|\Psi|^2+\frac{g}{2}|\Psi|^4\Big]d^3{\bf r}
\end{equation}
under additional condition of the presence of a given shape vortex, and with conservation of 
the total number of atoms assumed. From the minimum ${\cal H}$ requirement, in the limit $\xi\ll R_*$,
approximate conditions on the velocity field ${\bf v}=(\hbar/m_{\rm at})\nabla\Phi$ follow in the form
\begin{equation}
\mbox{div\,}(\rho({\bf r}){\bf v})=0, \qquad \mbox{curl\,}
{\bf v}=\Gamma\oint\delta({\bf r}-{\bf R}){\bf R}_\beta d\beta,
\label{velocity}
\end{equation}
where $\rho({\bf r})=m_{\rm at}|\Psi_0({\bf r})|^2\approx m_{\rm at}[\mu-V({\bf r})]/g$ is the 
undisturbed equilibrium condensate density in the absence of vortex, and
$\Gamma=2\pi\hbar/m_{\rm at}$ is the velocity circulation quantum.

It is very important for our purposes that from the Hamiltonian structure of the Gross-Pitaevskii 
equation governing the wave function,
\begin{equation}
i\hbar\Psi_t={\delta{\cal H}}/{\delta\Psi^*},
\label{Psi_variat}
\end{equation}
a variational equation follows for the filament motion at $\xi\ll R_*$,
\begin{equation}
\Gamma[{\bf R}_\beta\times{\bf R}_t]\rho({\bf R})\approx \delta{\cal H}/\delta{\bf R(\beta)}.
\label{variat_R}
\end{equation}
A direct proof of this statement for a uniform condensate can be found in a recent work \cite{BN2015}.
Generalization to the nonuniform case is very simple, so we do not place it here.
Earlier, Eq.(\ref{variat_R}) was derived in a more complicated way in Ref.\cite{R2001}.
It is the equation (\ref{variat_R}), which will serve as a basis of the theory developed below.
It will simplify significantly our consideration of vortex knot dynamics, since it will allow us, first,
to derive approximate equations of motion in the most compact and controlled manner; second, 
--- to use methods of Hamiltonian mechanics for their analysis.

A fully consistent application of the hydrodynamical approximation would require solution of the 
auxiliary equations (\ref{velocity}), and that would result in the Hamiltonian of a vortex line 
${\cal H}\{{\bf R}(\beta)\}$ as a double loop integral with involvement of a three-dimensional
matrix Green function (for technical details, see Ref.\cite{R2017-2}):
\begin{equation}
{\cal H}\{{\bf R}(\beta)\}=\frac{\Gamma^2}{2}\oint\oint 
{\bf R}_1'\cdot \hat G({\bf R}_1, {\bf R}_2) {\bf R}_2'd\beta_1 d\beta_2,
\end{equation}
where ${\bf R}'={\bf R}_\beta$. But if configuration of vortex line is far from self-intersections
(as, for example, in the case of a moderately deformed vortex ring) then the local induction 
approximation (LIA) is applicable,
\begin{equation}
{\cal H}\{{\bf R}(\beta)\}\approx {\cal H}_{LIA}=
\frac{\Gamma^2\Lambda}{4\pi}\oint\rho({\bf R})|{\bf R}'|d\beta,
\end{equation}
where $\Lambda=\ln(R_*/\xi)\approx$ const $\gg 1$ is a large logarithm. Substitution of this expression
into Eq.(\ref{variat_R}) and subsequent resolution with respect to the time derivative give 
the local induction equation \cite{SF2000,FS2001,R2001}
\begin{equation}
{\mathbf R}_t\big|_{\rm norm}=\frac{\Gamma \Lambda}{4\pi}\Big(\varkappa {\mathbf b}
+[\nabla\ln\rho({\mathbf R})\times {\boldsymbol \tau}]\Big),
\label{LIA}
\end{equation}
where  $\varkappa$ is a local curvature of the line, ${\mathbf b}$ is a unit bi-normal vector, 
and ${\boldsymbol \tau}$ is a unit tangent vector.

It is a well known fact that at $\rho =$ const, the local induction equation is reduced by the Hasimoto
transform to a one-dimensional focusing nonlinear Schrodinger equation \cite{Hasimoto}, so
the dynamics of a vortex filament at a uniform background can be nearly integrable. As to nonuniform
density profiles, examples of application of this model can be found in 
Refs.\cite{Kelvin_vaves,ring_istability,R2016-1,R2016-2,R2017-3}.

But in a number of interesting cases the LIA turns out to be certainly insufficient.
For instance, let in the system a closed vortex filament like a torus knot ${\cal T}_{P,Q}$ be present
(where $P>1$ and $Q>1$ are co-prime integers), and its shape in cylindrical coordinates be specified 
by two $2\pi P$-periodic on the angle $\varphi$ functions $Z(\varphi,t)$ and $R(\varphi,t)$.
Then it is easy to evaluate that with reasonable values of the parameter $\Lambda=5.0\div 9.0$, 
a contribution of the local induction into the filament dynamics is though essential, but sub-dominant
as compared to non-local quasi-two-dimensional interactions between different parts 
of the same vortex line corresponding to $\varphi$ values differing by $2\pi n$. 
Vortex knots on a uniform density background were extensively investigated 
earlier (see  \cite{RSB999,MABR2010,POB2012,KI2013,POB2014,LMB016,KKI2016} and references therein),
while analytical results for knots in nonuniform condensates are in fact absent so far.

The purpose of this work is to study the dynamics of simplest vortex knots in a trapped axisymmetric
Bose-Einstein condensate with equilibrium density profile $\rho(z,r)$. We will derive simplified
equations of motion for the case when function  $r\rho(z,r)$ has a quadratic maximum at a point 
$z=0$, $r=R_*$, so that
$$
\frac{\delta(r\rho)}{R_*\rho_*}\approx -\Big[(\alpha-\epsilon) \frac{z^2}{2R_*^2} 
+(\alpha+\epsilon) \frac{(\delta r)^2}{2R_*^2}\Big], \quad 0\le|\epsilon|<\alpha,
$$
and the shape of vortex filament is described by rather small functions $u=Z(\varphi,t)/R_*$
and $v=R(\varphi,t)/R_* -1$. Analytical solutions of the approximate equations will be found
corresponding to stationary rotating around $z$-axis vortex torus knots. In some domains of 
parameters $\alpha$ and $\epsilon$, the stationary solutions turn out to be unstable,
and that fact seemingly should correspond to long-lived knots in the original system described 
by the Gross-Pitaevskii equation.

\section{Derivation of simplified equations} 

To keep formulas clean, below we use dimensionless units, so that $\Gamma/{2\pi}=1$, $R_* = 1$, 
$\rho_*=1$. We divide the full period $2\pi P$ of functions $u$ and $v$ on the azimuthal angle into $P$ 
equal parts of length $2\pi$ each, and introduce notations $u_j(\varphi)\equiv u(2\pi j+\varphi)$, 
$v_j(\varphi)\equiv v(2\pi j+\varphi)$, where the index $j$ runs from $0$ to $P-1$. We will suppose
the following inequalities are valid,
$$
(u,v)\ll 1, \quad a_{jl}\equiv\sqrt{(u_j-u_l)^2+(v_j-v_l)^2}\gg(\xi/R_*).
$$
Apparently, typical values $\tilde u$, $\tilde v$, and $\tilde a_{jl}$ are all of the same order of magnitude.

Now we are going to use the fact that vector equation (\ref{variat_R}) with the chosen parametrization 
of vortex line is equivalent to non-canonical Hamiltonian system
\begin{eqnarray}
 (1+v)\rho(u,1+v) u_t&=&\delta H/\delta v,
 \label{u_t}\\
-(1+v)\rho(u,1+v) v_t&=&\delta H/\delta u.
 \label{v_t}
\end{eqnarray}
Since at small $u$ and $v$ we have
\begin{equation}
(1+v)\rho(u,1+v)\approx 1-(\alpha-\epsilon) u^2/2 -(\alpha+\epsilon) v^2/2,
\end{equation}
it will be sufficient to put in the l. h. sides of Eqs.(\ref{u_t})-(\ref{v_t}) $(1+v)\rho(u,1+v)\approx 1$. 
In fact, that will mean neglecting terms of order $\tilde u^2/\tilde a_{jl}\sim \tilde u$ in comparison
to terms of order $\Lambda \tilde u$ in equations of motion, as it will become clear from further 
consideration. As to the r.h. sides,  we will use an approximate Hamiltonian $H\approx H_{LIA}^{(2)}+H_{2D}$,
with the first part being the expanded to the second order Hamiltonian of local induction in terms of
functions $u$ and $v$, while the second part is the interaction Hamiltonian between strictly co-axial
perfect vortex rings near the maximum of function $r\rho$:
\begin{equation}
H_{LIA}^{(2)}=\frac{\Lambda}{2}\int\limits_0^{2\pi P}\Big[\frac{u_\varphi^2}{2}+\frac{v_\varphi^2}{2}
-(\alpha-\epsilon) \frac{u^2}{2} -(\alpha+\epsilon) \frac{v^2}{2} \Big]d\varphi,
\end{equation}
\begin{equation}
H_{2D}\approx\frac{1}{2}\int\limits_0^{2\pi}\sum_j\sum_{l\neq j}
\ln\Big[(u_j-u_l)^2 +(v_j-v_l)^2\Big]^{-\frac{1}{2}} d\varphi,
\label{inter}
\end{equation}
and summation on $l$ and $j$ in the double sum goes from $0$ to $P-1$, excluding diagonal terms.
Generally speaking, instead of the logarithm in Eq.(\ref{inter}),
there should appear a Green function $G$ which is a solution of equation
\begin{equation}
-\partial_z\Big[\frac{G_z}{r\rho(z,r)}\Big]-\partial_r\Big[\frac{G_r}{r\rho(z,r)}\Big]
=2\pi\delta(z-z_0)\delta(r-r_0),
\end{equation}
and actually it is a cylindrical stream function created by a point vortex [placed at position $(z_0,r_0)$]
in the half-plane  $(z,r)$ for divergence-free field $\rho{\bf v}$. Formula  
$$
G(z,r;z_0,r_0)=\sqrt{r\rho(z,r)r_0\rho(z_0,r_0)}\tilde G(z,r;z_0,r_0),
$$ 
is valid, with function $\tilde G$ satisfying the equation 
\begin{equation}
[-\partial^2_z-\partial^2_r +\tilde\kappa^2(z,r)]\tilde G=2\pi\delta(z-z_0)\delta(r-r_0),
\end{equation}
where $\tilde\kappa^2(z,r)=\sqrt{r\rho}[\partial^2_z+\partial^2_r](1/\sqrt{r\rho})$. In the vicinity 
of the maximum we have $\tilde\kappa^2\approx\alpha+{\cal O}(u^2,v^2)$, and therefore with sufficient
accuracy 
$$
\tilde G\approx K_0\Big(\sqrt{\alpha}\sqrt{(u-u_0)^2+(v-v_0)^2}\Big),
$$ 
where $K_0(\dots)$ is the modified Bessel function. Thus, 
\begin{eqnarray}
G&\approx&\Big[1-\frac{1}{4}(\alpha-\epsilon)(u^2+u_0^2)-\frac{1}{4}(\alpha+\epsilon)(v^2+v_0^2)\Big]
\nonumber\\
&&\qquad \times K_0\Big(\sqrt{\alpha}\sqrt{(u-u_0)^2+(v-v_0)^2}\Big).
\end{eqnarray}
Replacing this expression by the logarithm in the Hamiltonian, we neglect terms of order 
$\tilde u^2\ln \tilde u$ in comparison with the main contribution from local induction which 
has the order $\Lambda \tilde u^2$. It is also clear we have neglected those effects in 
the interaction Hamiltonian which are related to a difference between actual filament shape and 
a perfect co-axial ring.

As the result of all the simplifications made, a canonical Hamiltonian system of nonlinear equations
on $u_n$ and $v_n$ is obtained, which is able to describe dynamics of vortex knots:
\begin{eqnarray}
u_t&=&-\frac{\Lambda}{2}[v_{\varphi\varphi}+(\alpha+\epsilon)v]\nonumber\\
&&\qquad-\sum_{j\neq n}\frac{(v_n-v_j)}{(u_n-u_j)^2+(v_n-v_j)^2},\\
-v_t&=&-\frac{\Lambda}{2}[u_{\varphi\varphi}+(\alpha-\epsilon)u]\nonumber\\
&&\qquad-\sum_{j\neq n}\frac{(u_n-u_j)}{(u_n-u_j)^2+(v_n-v_j)^2}.
\end{eqnarray}
Re-scaling here the time as $\Lambda t_{\rm old}= t_{\rm new}$, and introducing complex-valued functions
$W_n=\sqrt{\Lambda}(u_n+iv_n)$, we represent our system in a more compact form,
\begin{equation}
iW_{n,t}=-\frac{1}{2}(W_{n,\varphi\varphi}+\alpha W_n-\epsilon W^*_n)
-\sum_{j\neq n}\frac{1}{(W^*_n-W^*_j)},
\label{W_t}
\end{equation}
with the cyclic permutation in the boundary conditions, $W_0(2\pi)=W_1(0)$, $W_1(2\pi)=W_2(0)$, 
$\dots$, $W_{P-1}(2\pi)=W_0(0)$. Let us note on the way, the same system but with different 
boundary conditions, when the permutation contains several cycles,
is able to describe dynamics of several vortex filaments including linked and knotted ones.

We also note that sum $w=\sum_n W_n$ satisfies the linear equation
\begin{equation}
iw_t=-\frac{1}{2}(w_{\varphi\varphi}+\alpha w-\epsilon w^*)
\label{w}
\end{equation}
with periodic boundary conditions. It is easy for solution and has the eigen-frequencies 
\begin{equation}
\omega_m=\mbox{sgn}(m^2-\alpha+|\epsilon|) \sqrt{(m^2-\alpha+\epsilon)(m^2-\alpha-\epsilon)},
\end{equation}
where $m$ is an integer. Taking into account next-order corrections on amplitudes would result in
considerably more cumbersome equations of motion than Eq.(\ref{W_t}). In many cases such corrections 
are not necessary. Exceptions are possible parametric resonances as $\omega_0=-2\omega_m$, which 
require special relations between $\alpha$ and $\epsilon$ (see examples in Ref.\cite{R2017-3}).
We assume here non-resonant situation for simplicity.

\section{Solutions at $\epsilon =0$}

A spatial inhomogeneity of the condensate came into Eqs.(\ref{W_t}) through the coefficients 
$\alpha$ and $\epsilon$. They both have essential influence on the dynamics only when $\epsilon\neq 0$.
If $\epsilon=0$ then a simple phase factor as $W_n=\Theta_n\exp(i\alpha t/2)$ removes the terms with 
$\alpha$ from the system, resulting in well-known equations which are used for modeling long-wave 
dynamics of weakly curved, nearly parallel vortex filaments in a uniform perfect fluid (see, e.g.,
\cite{v_filaments1,v_filaments2,v_filaments3,v_filaments4}, and references therein):
\begin{equation}
i\Theta_{n,t}=-\frac{1}{2}\Theta_{n,\varphi\varphi} -\sum_{j\neq n}\frac{1}{(\Theta^*_n-\Theta^*_j)}.
\label{Theta_t}
\end{equation}
In this case, besides the corresponding Hamiltonian
\begin{equation}
H_0=\int\limits_0^{2\pi}\Big(\sum_n |\Theta_{n,\varphi}|^2/2 
-\sum_n\sum_{l\neq n}\ln|\Theta_n-\Theta_l|\Big) d\varphi
\label{H_0}
\end{equation}
and the angular momentum (with respect to $z$-axis)
\begin{equation}
M=\frac{1}{2i}\int_0^{2\pi}\sum_n(\Theta_n^*\Theta_{n,\varphi}-\Theta_n\Theta^*_{n,\varphi})d\varphi,
\end{equation}
there are additional integrals of motion in the system,
\begin{equation}
N_{\sigma}=\int_0^{2\pi}\sum_n |\Theta_n-\sigma|^2 d\varphi.
\end{equation}
Consequently, stable (within the approximate model under consideration) solutions exist in the form
$\Theta_n=\sigma+ \theta_n(\varphi-\gamma t)\exp(i\lambda t)$, each configuration corresponding to 
a local minimum of a bounded from below functional $\tilde H_0= H_0+\lambda N_\sigma -\gamma M$,
with some constants $\lambda$, $\sigma$, and $\gamma$ (without loss of generality, we put $\sigma=0$).
In particular torus knots ${\cal T}_{P,Q}$, when
\begin{equation}
\theta_n=c\exp\Big(i\frac{2\pi nQ}{P}+i\frac{Q}{P}\varphi\Big), 
\end{equation}
are certainly stable at $P\leq 6$, which property follows from the classical result about stability
of regular $P$-polygon of point vortices on a plane.

Thus, we have obtained an interesting theoretical result: at definite parameters of external
potential, long-lived vortex knots can exist in the condensate. In particular, for an anisotropic 
harmonic trap, where the density profile in the Thomas-Fermi regime is
$$
\rho=3/2-(r^2+\alpha_\parallel z^2)/2,
$$ 
the coefficients of system (\ref{W_t})  are
$$
\alpha=(3+\alpha_\parallel)/2, \qquad \epsilon=(3-\alpha_\parallel)/2.
$$
To have the most stable knots, it is necessary to use here the anisotropy parameter 
$\alpha_\parallel\approx3$.

\begin{figure}
\begin{center}
\epsfig{file=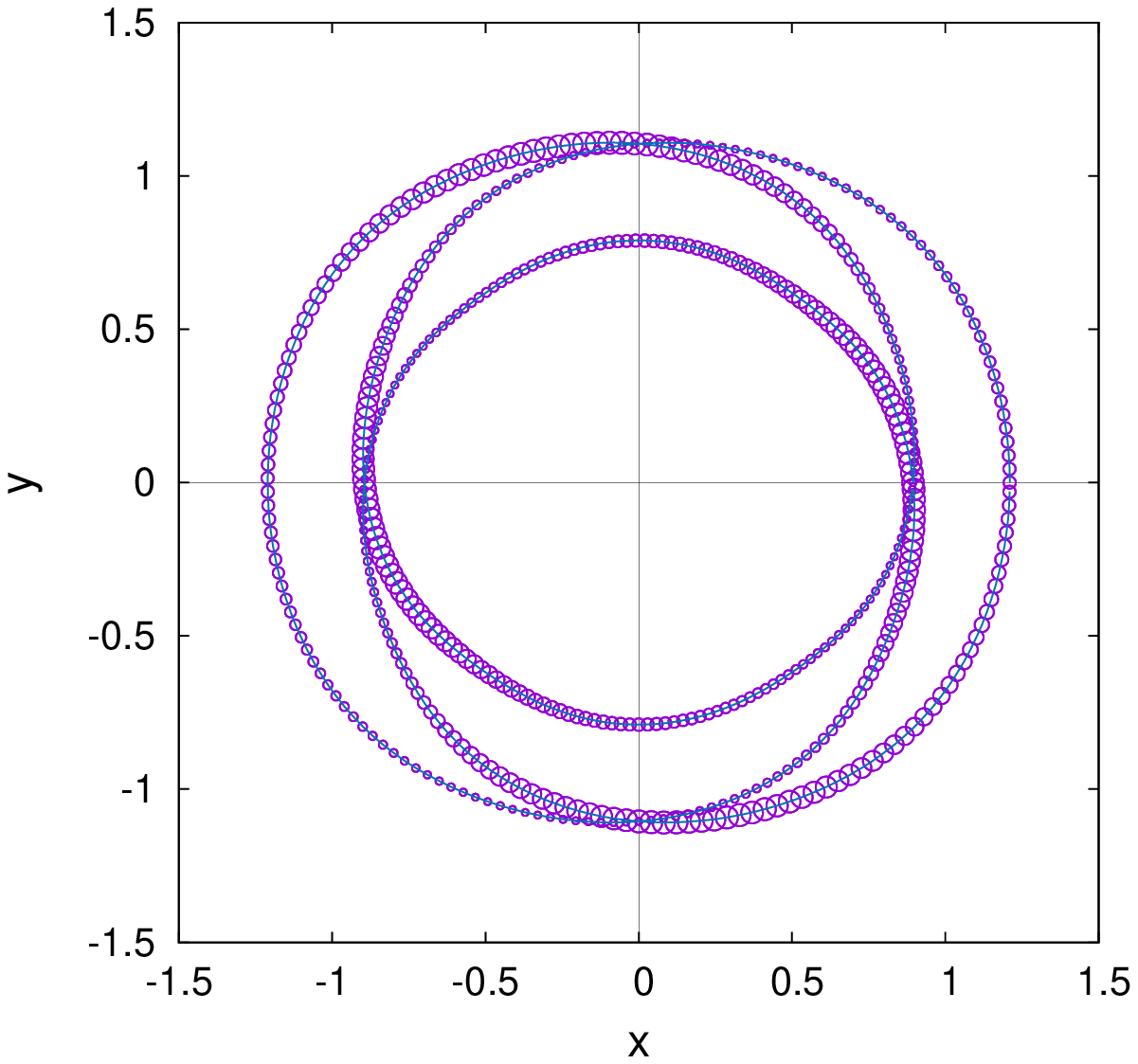, width=69mm}\\
\epsfig{file=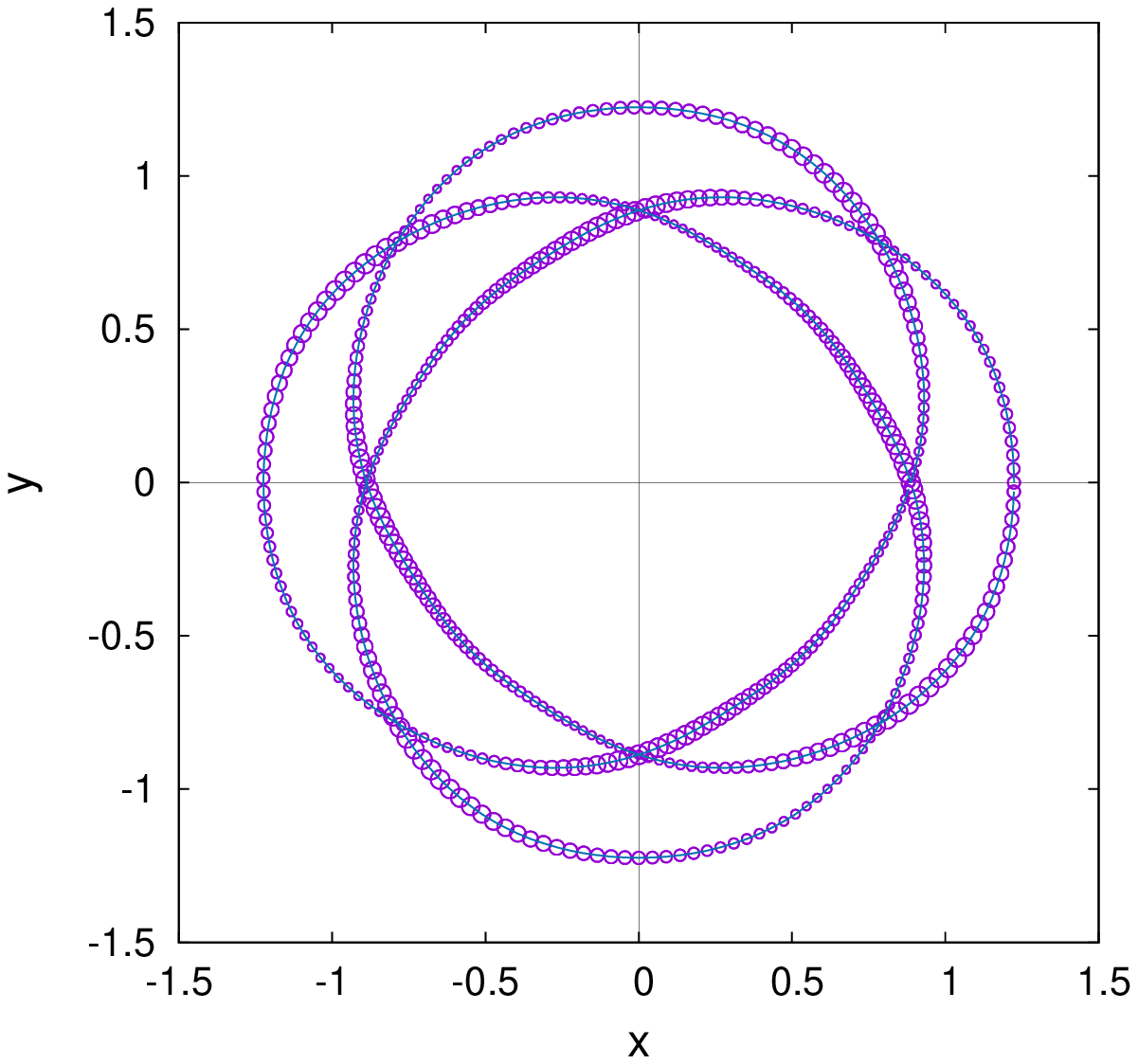, width=69mm}\\
\epsfig{file=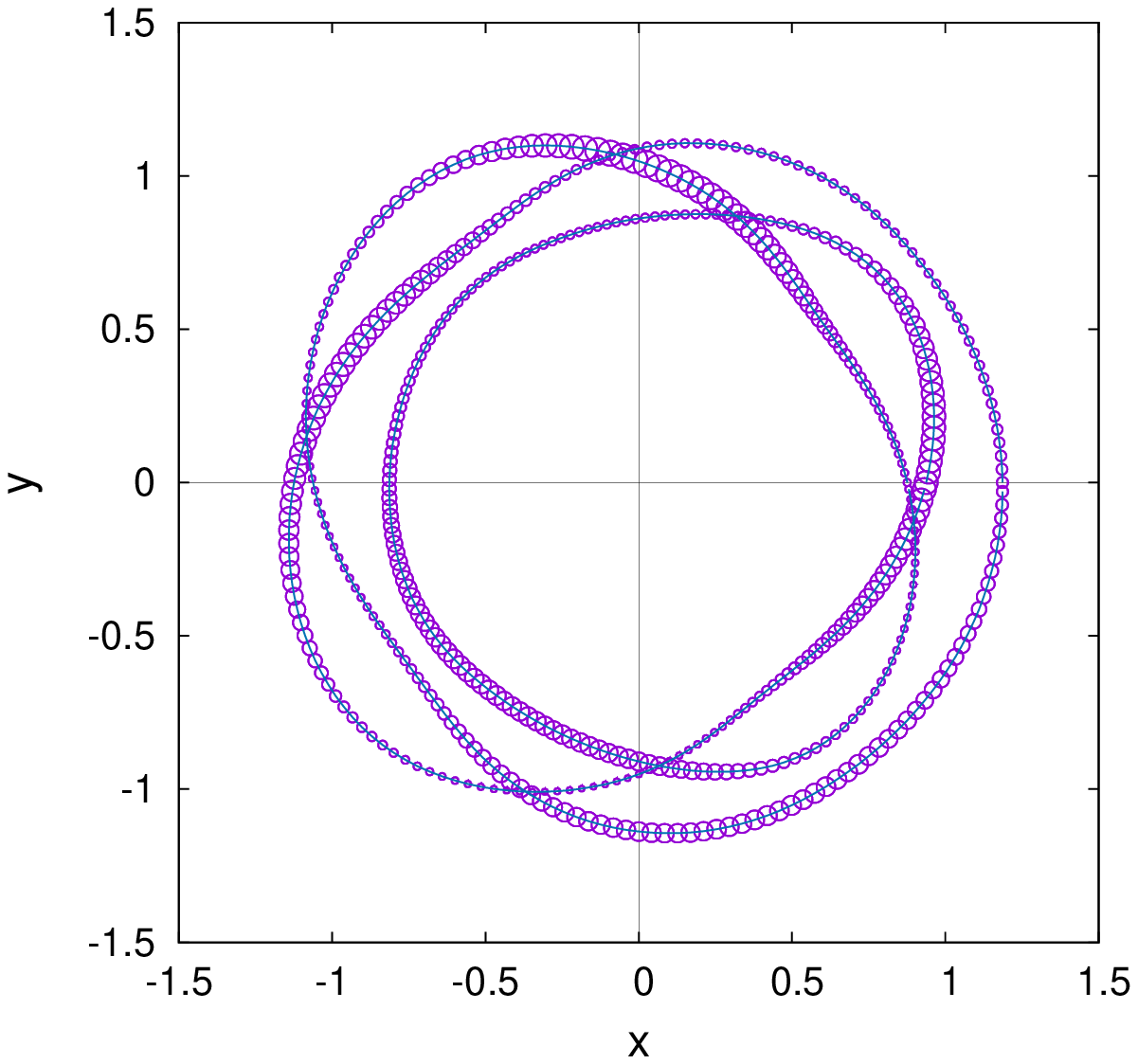, width=69mm}
\end{center}
\caption{Examples of stable knots at $P=3$, found by simulation of equations  (\ref{gradient})
with different initial conditions:
a) $f(\varphi,0)=0.6[i\cos(2\varphi/3)-0.5\sin(4\varphi/3)]$ --- ``figure-eight knot''; 
b) $f(\varphi,0)=0.6[i\cos(4\varphi/3)-0.5\sin(8\varphi/3)]$; c) 
$f(\varphi,0)=0.6i\{\cos(\varphi/3)+\exp(11i\varphi/3)[1+\cos(2[\varphi-\pi]/3)]^4/16\}$ ---
``granny knot''. In cases a) and b) the parameters $\lambda=4.0$,  $\gamma=0$. 
In case c) $\gamma=1.5$, $\lambda=\gamma^2/2+4.5$.}
\label{braid} 
\end{figure}

Numerical finding of examples of such ``stationary'' solutions is easy and can be efficiently made
by the method of gradient descent, that is by simulation of an auxiliary evolutionary system
\begin{equation}
f_{n,\tau}=-\frac{\delta \tilde H_0}{\delta f^*_n}=\frac{1}{2}f_{n,\varphi\varphi}
-i\gamma f_{n,\varphi}-\lambda f_n
+\sum_{j\neq n}\frac{1}{(f^*_n-f^*_j)}.
\label{gradient}
\end{equation}
Several nontrivial examples for $P=3$ are shown in Fig.\ref{braid} (with the choice $\Lambda =6.25$;
the line in the figure is more thick where a local value of coordinate $z$ is larger).

But it would be a mistake to believe that, having taken arbitrary initial data $f_n(\varphi,\tau=0)$, 
we can integrate the gradient equations to very long $\tau$ with conservation of initial topology
of the knot, and that in the limit $\tau\to+\infty$ we will inevitably get some smooth functions 
$f_n\to\theta_n$. On the contrary, it is quite possible that after a finite ``time'',
a singularity will develop in system  (\ref{gradient}), when 
$\delta(\tau)\equiv\mbox{min}_\varphi |f_{j}(\varphi, \tau)-f_{l}(\varphi, \tau)|\to 0$ 
at $\tau\to \tau_{\rm sing}$. Numerical discretization of the curve will result in a self-intersection,
and topology will change at that event. In other words, a dispersive tendency towards smoothing
of locally twisted filaments can prevail over their mutual repulsion. Indeed, when a typical scale
along $\varphi$ becomes of order $\delta$, then the dispersive term $f_{\varphi\varphi}$ has the same 
order $1/\delta$ as the nonlinear term. Therefore the question remains open, if a stationary solution
always exists at a given knot topology.
 
But nevertheless, in many cases $\delta(\tau)$ does not go to zero, topology of initial knot
is conserved, and smooth solutions are found successfully (sometimes, however, the ``relaxation'' 
to a minimum of $\tilde H_0$ occurs rather slowly).

We also mention a possible case $\alpha<0$, when function $r\rho(z,r)$ has a minimum at point
$z=0$, $r=1$ instead of maximum. In that case at $\epsilon\neq 0$  the integrals $N_\sigma$ are
absent, but the functional $H-\gamma M$ itself is bounded from below. Therefore stable rotating 
around $z$-axis knots of the form  $W_n=F_n(\varphi-\gamma t)$ exist.

\begin{figure}
\begin{center}
\epsfig{file=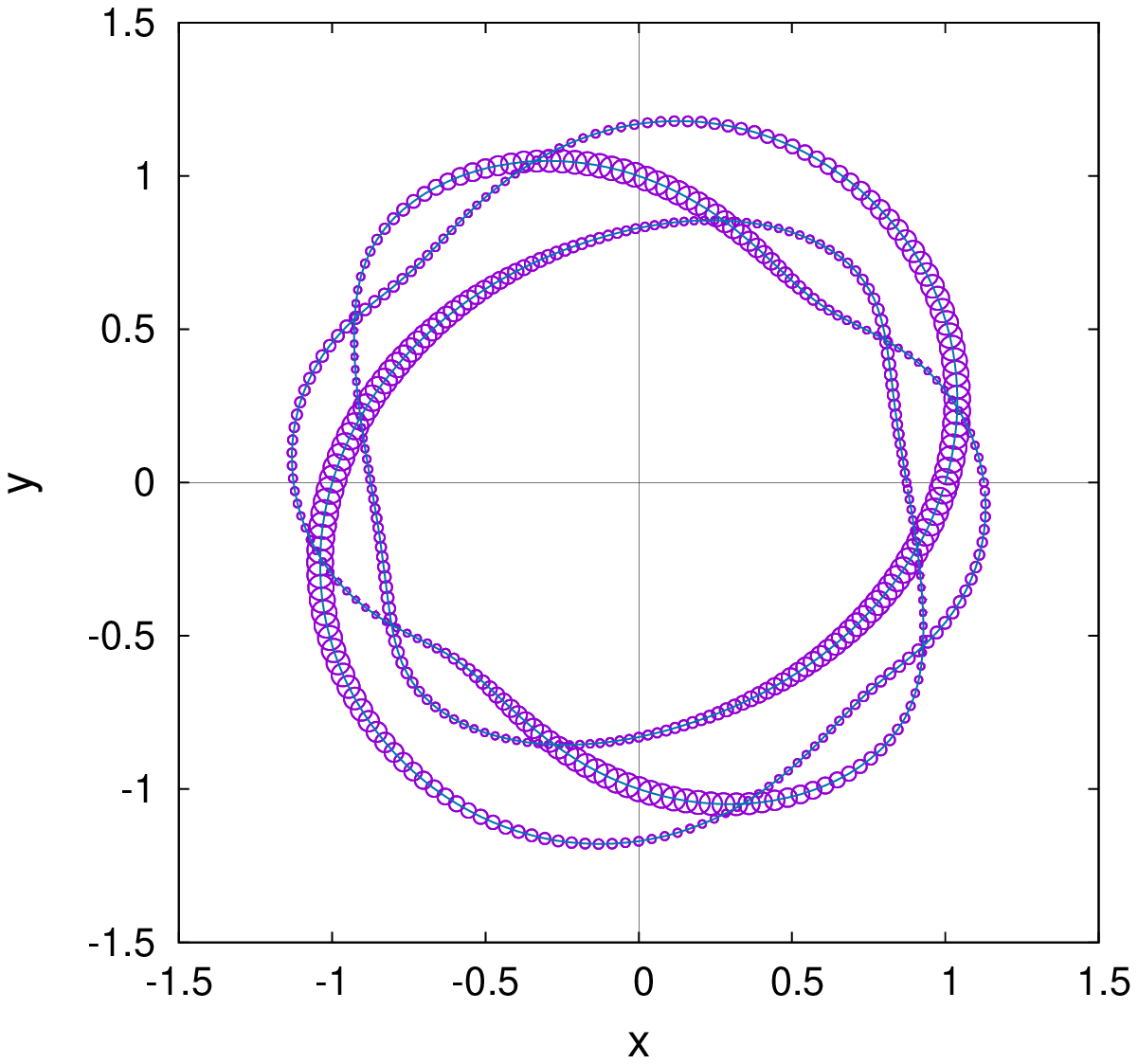, width=75mm}\\
\epsfig{file=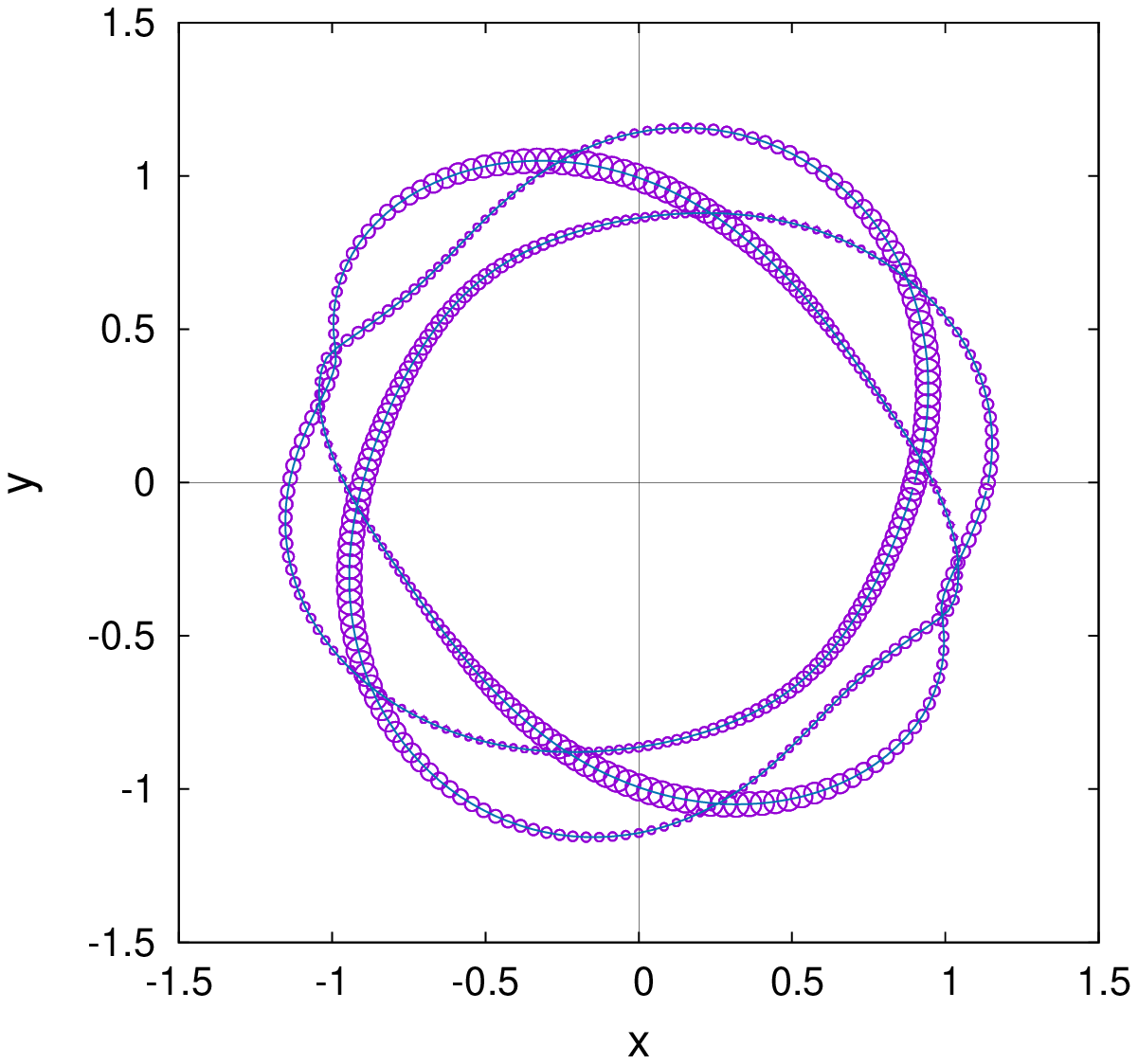, width=75mm}
\end{center}
\caption{A tendency towards finite-time singularity in Eqs.(\ref{Theta_t}): 
a) initial knot configuration obtained by simulation of auxiliary system (\ref{gradient}) 
(with parameters $\gamma=0.7$, $\lambda=\gamma^2/2+4.0$) to $\tau_{\rm max}=7.0$, with initial conditions 
$f(\varphi,0)=0.6[\exp(2i\varphi/3)+ \exp(4i\varphi/3)+\exp(8i\varphi/3)]$; b) the knot at $t=2.3$
with locally convergent parts of the filament [shown without taking into account the phase factor 
$\exp(i\alpha t/2)$].}
\label{FTS} 
\end{figure}

Another noteworthy result is about numerical simulations of equations of motion.
It turns out that also conservative system (\ref{Theta_t}) can exhibit a tendency towards 
a finite-time  singularity, not only the gradient system (\ref{gradient}). An appropriate example 
is given in Fig.\ref{FTS}. In the framework of the Gross-Pitaevskii equation, such a rapid collision
of vortex lines typically results in their re-connection. Of course, our simplified equations loose
their literal applicability closely to the singularity, but an initial stage of its formation is 
described correctly. Clarification of collapse conditions and its relation to vortex knot 
topology are interesting tasks for future research.

\begin{figure}
\begin{center}
\epsfig{file=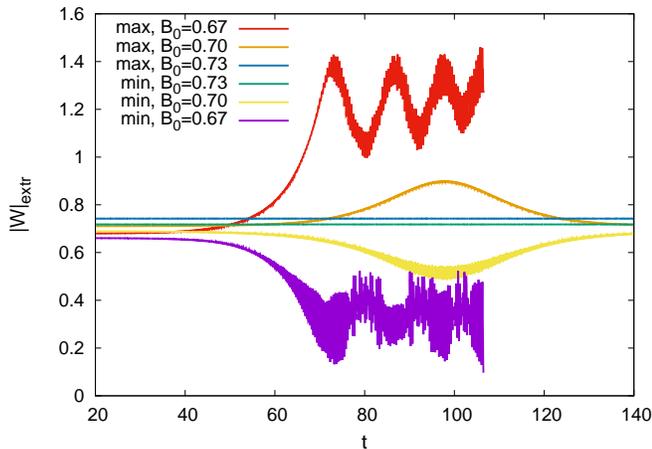, width=88mm}
\end{center}
\caption{Examples of the dynamics of extreme deviations for stable and unstable solutions (\ref{stat_W}).}
\label{W_max_min_t} 
\end{figure}
\begin{figure}
\begin{center}
\epsfig{file=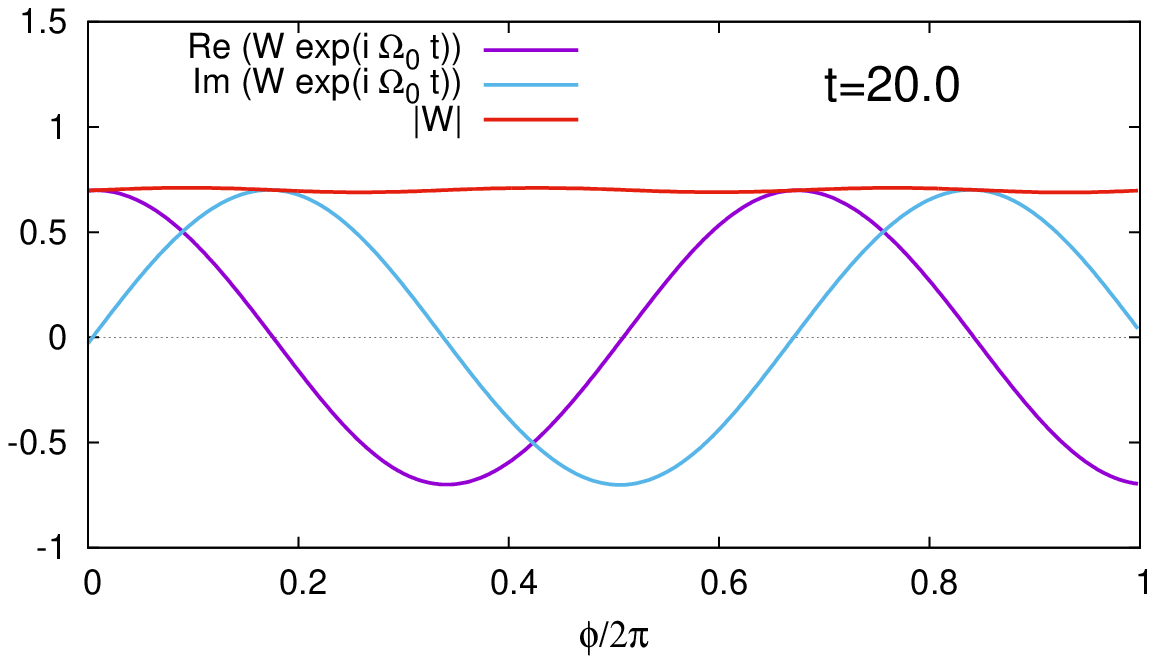, width=80mm}\\
\epsfig{file=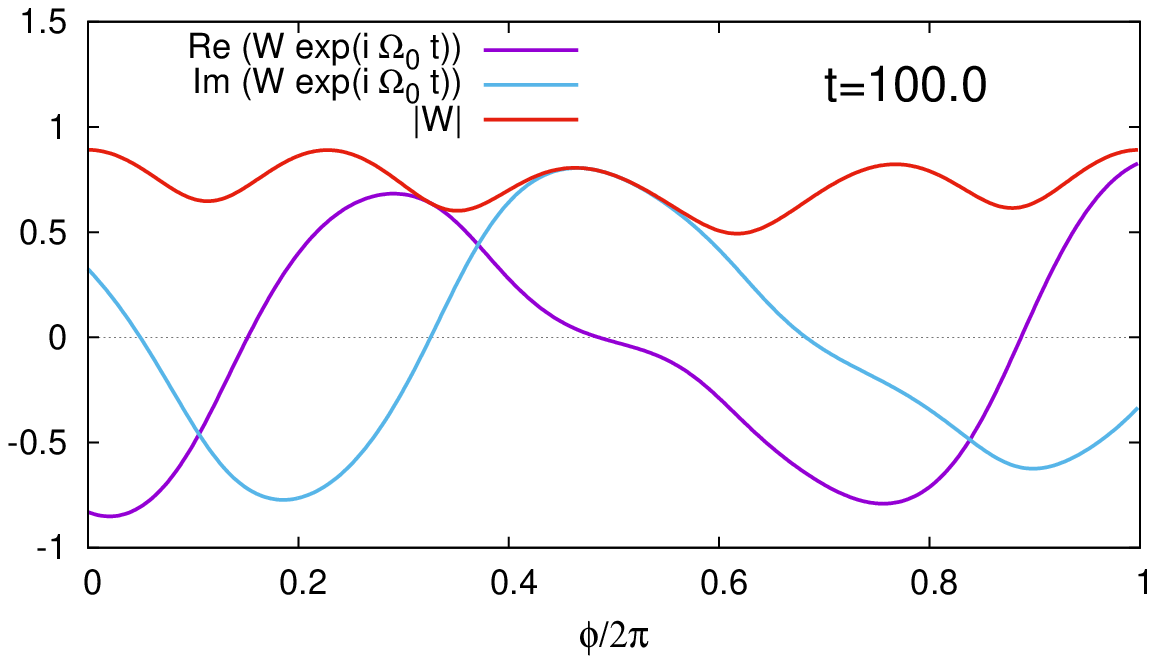, width=80mm}\\
\epsfig{file=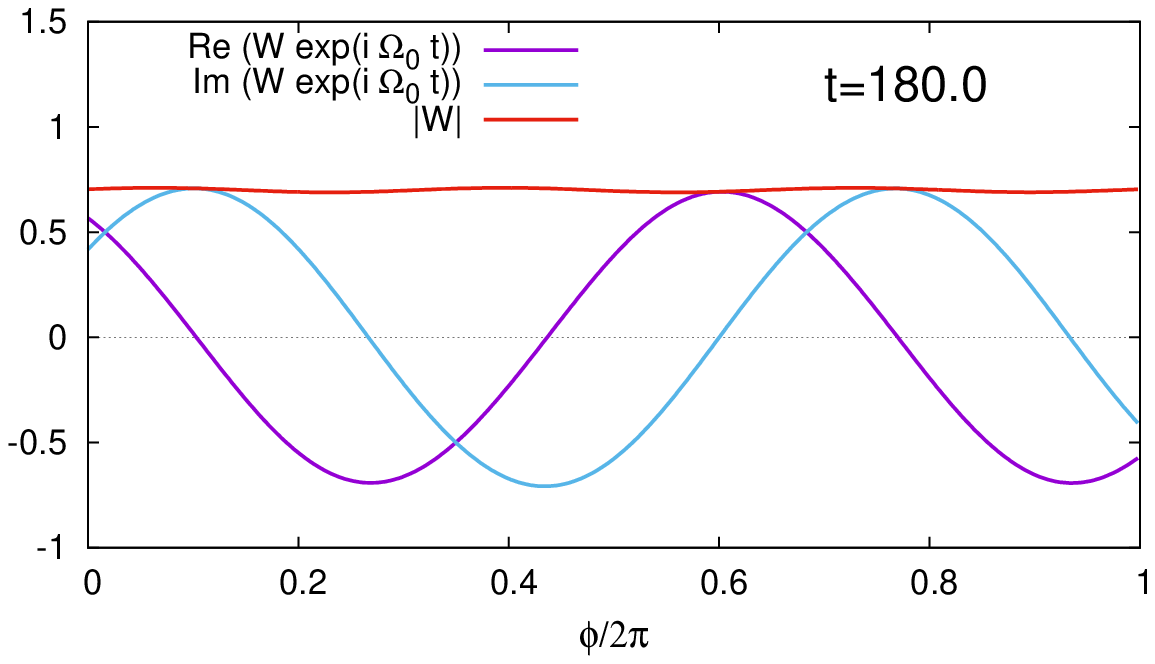, width=80mm}
\end{center}
\caption{Growth and subsequent decay of unstable mode at  $\alpha=2.8$, $\epsilon=0.2$, $B_0=0.7$.}
\label{W_123} 
\end{figure}
\begin{figure}
\begin{center}
\epsfig{file=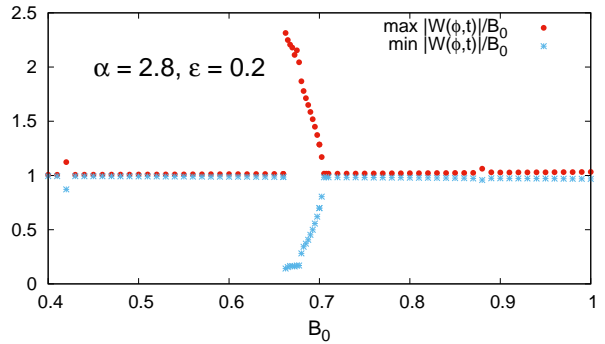, width=88mm}
\end{center}
\caption{Structure of stable and unstable bands at $\alpha=2.8$, $\epsilon=0.2$. One strongly unstable 
band is seen near $B_0=0.66$, and two weakly unstable bands are near $B_0=0.42$ and $B_0=0.88$.}
\label{W_MAX_MIN_B0} 
\end{figure}

\section {Case $P=2$}

Let us now consider in more detail the simplest vortex configurations with $P=2$.
Introducing the sum $w=W_0+W_1$ and the difference  $W=W_0-W_1$, we obtain two ``un-linked'' equations:
Eq.(\ref{w}) and
\begin{equation}
iW_t=-\frac{1}{2}(W_{\varphi\varphi}+\alpha W-\epsilon W^*)-\frac{2}{W^*},
\end{equation}
the last one with anti-periodic boundary conditions. Let us concentrate our attention on this equation.
In the case $\epsilon=0$, exact solutions exist of the form
\begin{equation}
W=B_0\exp(ik_0\varphi -i\Omega_0 t +i\zeta_0), 
\end{equation}
depending on a real parameter $B_0>0$, where 
\begin{equation}
\Omega_0=\frac{1}{2}(k_0^2-\alpha)-\frac{2}{B_0^2},
\end{equation}
and $k_0$ should be a half-integer $Q/2$ (at $k_0=3/2$ the solution corresponds to ``trefoil''
${\cal T}_{2,3}$). It is not difficult to analyze small perturbations and make sure that 
the given solution is stable at all $B_0$ and $\alpha$.

Situation is more complicated when  $\epsilon\neq 0$. In that case there are stationary solutions 
as well. At small $\epsilon$ up to the first-order accuracy we easily obtain
\begin{equation}
W\approx B_0\exp(i\zeta)+\epsilon C\exp(-i\zeta) +\epsilon D\exp(3i\zeta),
\label{stat_W}
\end{equation}
where $\zeta=k_0\varphi -\Omega_0 t +\zeta_0$, and coefficients $C$ and $D$ 
satisfy a linear system of equations
\begin{eqnarray}
-\Omega_0 C&=&\frac{1}{2}(k_0^2-\alpha)C +\frac{2}{B_0^2} D+\frac{B_0}{2},\\
3\Omega_0 D&=&\frac{1}{2}(9k_0^2-\alpha)D +\frac{2}{B_0^2} C.
\end{eqnarray}
The solution of this system is
\begin{equation}
D=B_0/J, \qquad C=-B_0^3(3k_0^2+\alpha+6/B_0^2)/(2J), 
\end{equation}
\begin{equation}
J=-4/B_0^2+B_0^2(k_0^2-\alpha-2/B_0^2)(3k_0^2+\alpha+6/B_0^2).
\end{equation}
A difference in comparison to the case $\epsilon=0$ is that at some relations between $B_0$ and $\alpha$,
parametric resonances turn out be possible in the dynamical system under consideration, thus resulting in 
instabilities. In other words, with small random initial perturbations near the stationary solution, 
three possible scenarios can develop in subsequent dynamics of the largest and the smallest
values of  $|W|$. When solution (\ref{stat_W}) is stable, $|W|_{\rm max,min}$ do not exhibit considerable 
changes. But if  (\ref{stat_W}) is unstable then deviations of $|W|$ first grow exponentially, and after 
that either return closely to initial small values or the system passes to a quasi-turbulent regime.
For illustration, in Fig.\ref{W_max_min_t} examples of numerically found temporal dependencies of
extreme quantities $\mbox{max}_\varphi|W(\varphi,t)|$ and $\mbox{min}_\varphi|W(\varphi,t)|$ are given
at $\alpha=2.8$, $\epsilon=0.2$ for several $B_0$.
Fig.\ref{W_123} shows how the solution looks at recurrent dynamics of unstable mode.

If in each such numerical simulation on a sufficiently long time interval we record globally reached
extreme values of quantity $|W|/B_0$, and after that we draw their dependencies on $B_0$, then a structure
of stable and unstable bands at fixed $\alpha$ and $\epsilon$ is obtained. An example is given
in Fig.\ref{W_MAX_MIN_B0}. It is important that with rather small $\epsilon$, stable bands dominate.

\section{Conclusions}

Thus, in this work for the first time relatively simple and convenient for analytical study and 
for numerical modeling, Hamiltonian equations have been derived which are able to describe in 
a definite limit the dynamics of knotted quantum vortex lines in spatially nonuniform axisymmetric 
Bose-Einstein condensates in the Thomas-Fermi regime at zero temperature. In numerical experiments,
in a number of cases, a tendency towards a finite-time singularity was observed for knots with 
nontrivial topology. In the framework of suggested model, at definite values of parameters, stable
stationary solutions with different topology have been found, with no analogs in uniform condensates.
It has been revealed that in general case, solutions in the form of torus knots can be parametrically
unstable but unstable bands are relatively narrow at small anisotropy of the maximum of function
$r\rho(z,r)$, in the vicinity of which all the dynamics occurs.

Our theory is heavily based upon a large parameter $\Lambda$ which however can hardly be very large 
in reality. Therefore, it seems very desirable to compare in the future the predictions of the present 
simplified model to results of direct numerical simulation of the Gross-Pitaevskii equation at 
moderate $\Lambda$.


\vspace{4mm}

\begin{thebibliography}{99}

\bibitem{F2009}  A. L. Fetter, Rev. Mod. Phys. {\bf 81}, 647 (2009).

\bibitem{SF2000} A. A. Svidzinsky and A. L. Fetter,
Phys. Rev. A {\bf 62}, 063617 (2000).

\bibitem{FS2001} A. L. Fetter and A. A. Svidzinsky,
J. Phys.: Condens. Matter {\bf 13}, R135 (2001).

\bibitem{R2001} V. P. Ruban, Phys. Rev. E {\bf 64}, 036305 (2001).

\bibitem{AR2001} A. Aftalion and T. Riviere, Phys. Rev. A {\bf 64}, 043611 (2001).

\bibitem{GP2001} J. Garcia-Ripoll and V. Perez-Garcia,
Phys. Rev. A {\bf 64}, 053611 (2001).

\bibitem{A2002} J. R. Anglin, Phys. Rev. A {\bf 65}, 063611 (2002).

\bibitem{AR2002} A. Aftalion and R. L. Jerrard, Phys. Rev. A {\bf 66}, 023611 (2002).

\bibitem{RBD2002} P. Rosenbusch, V. Bretin, and J. Dalibard,
Phys. Rev. Lett. {\bf 89}, 200403 (2002).

\bibitem{AD2003} A. Aftalion and I. Danaila, Phys. Rev. A {\bf 68}, 023603 (2003).

\bibitem{AD2004} A. Aftalion and I. Danaila, Phys. Rev. A {\bf 69}, 033608 (2004).

\bibitem{SR2004} D. E. Sheehy and L. Radzihovsky, Phys. Rev. A {\bf 70}, 063620 (2004).

\bibitem{D2005} I. Danaila, Phys. Rev. A {\bf 72}, 013605 (2005).

\bibitem{Kelvin_vaves} A. Fetter, Phys. Rev. A {\bf 69}, 043617 (2004).

\bibitem{ring_istability} T.-L. Horng, S.-C. Gou, and T.-C. Lin,
Phys. Rev. A {\bf 74}, 041603 (2006).

\bibitem{v-2015} S. Serafini, M. Barbiero, M. Debortoli, S. Donadello, F. Larcher, F. Dalfovo, 
G. Lamporesi, and G. Ferrari, Phys. Rev. Lett. {\bf 115}, 170402 (2015).

\bibitem{reconn-2017}
S. Serafini, L. Galantucci, E. Iseni, T. Bienaime, R. N. Bisset, C. F. Barenghi, F. Dalfovo, 
G. Lamporesi, and G. Ferrari, Phys. Rev. X {\bf 7}, 021031 (2017).

\bibitem{top-2017} R. N. Bisset, S. Serafini, E. Iseni, M. Barbiero, T. Bienaime, G. Lamporesi, 
G. Ferrari, and F. Dalfovo, Phys. Rev. A {\bf 96}, 053605 (2017).

\bibitem{RSB999} R. L. Ricca, D. C. Samuels, and C. F. Barenghi, J. Fluid Mech. {\bf 391}, 29 (1999).

\bibitem{MABR2010} F. Maggioni, S. Alamri, C. F. Barenghi, and R. L. Ricca, 
Phys. Rev. E {\bf 82}, 026309 (2010).

\bibitem{POB2012} D. Proment, M. Onorato, and C. F. Barenghi, 
Phys. Rev. E {\bf 85}, 036306 (2012).

\bibitem{KI2013} D. Kleckner and W. T. M. Irvine, Nature Physics {\bf 9}, 253 (2013). 

\bibitem{POB2014} D. Proment, M. Onorato, and C. F. Barenghi, 
J. Phys.: Conf. Ser. {\bf 544}, 012022, (2014).

\bibitem{LMB016} P. Clark di Leoni, P. D. Mininni, and M. E. Brachet, 
Phys. Rev. A {\bf 94}, 043605 (2016).

\bibitem{KKI2016} D. Kleckner, L. H. Kauffman, and W. T. M. Irvine, 
Nature  Physics {\bf 12}, 650 (2016).

\bibitem{BN2015} M. D. Bustamante and S. Nazarenko, Phys. Rev. E {\bf 92}, 053019 (2015).

\bibitem{R2017-1} V. P. Ruban, JETP Letters {\bf 105}, 458 (2017).

\bibitem{R2017-2} V. P. Ruban, JETP {\bf 124}, 932 (2017).

\bibitem{Hasimoto} H. Hasimoto, J. Fluid Mech. {\bf 51}, 477 (1972).

\bibitem{R2016-1} V. P. Ruban, JETP Letters {\bf 103}, 780 (2016).

\bibitem{R2016-2} V. P. Ruban, JETP Letters {\bf 104}, 868 (2016).

\bibitem{R2017-3} V. P. Ruban, JETP Letters {\bf 106}, 223 (2017).

\bibitem{v_filaments1} V. E. Zakharov, Usp. Fiz. Nauk {\bf 155}, 529 (1988).

\bibitem{v_filaments2} R. Klein, A. J. Majda,  and K. Damodaran, J. Fluid Mech. {\bf 288}, 201 (1995).

\bibitem{v_filaments3}  C. E. Kenig, G. Ponce, and L. Vega, Commun. Math. Phys. {\bf 243}, 471 (2003).

\bibitem{v_filaments4} N. Hietala, R. Hanninen, H. Salman, and C. F. Barenghi, 
Phys. Rev. Fluids {\bf 1}, 084501 (2016).

\end{thebibliography}
\end{document}